\documentclass[epj]{webofc}
\usepackage[varg]{txfonts}   

\newcommand{\be}{\begin{equation}}
\newcommand{\ee}{\end{equation}}
\newcommand{\ba}{\begin{eqnarray}}
\newcommand{\ea}{\end{eqnarray}}
\newcommand{\ban}{\begin{eqnarray*}}
\newcommand{\ean}{\end{eqnarray*}}
\newcommand \nn {\nonumber}
\woctitle{International Conference on New Frontiers in Physics 2013}
\begin{document}

\title{Energy loss in unstable QGP -  problem of the upper cut-off}

\author{Margaret E. Carrington
\inst{1}\fnsep\thanks{\email{Carrington@BrandonU.CA}} 
\and
Katarzyna Deja \inst{2}\fnsep\thanks{\email{kdeja@fuw.edu.pl}} \and
Stanis\l aw Mr\' owczy\' nski\inst{2,3}\fnsep\thanks{\email{mrow@fuw.edu.pl}}   
}

\institute{Department of Physics, Brandon University, Brandon, Manitoba, Canada
\and
National Centre for Nuclear Research, Warsaw, Poland
\and
Institute of Physics, Jan Kochanowski University, Kielce, Poland}

\abstract{The energy loss of a highly energetic parton in a weakly coupled quark-gluon plasma is studied as an initial value problem.  An extremely prolate plasma, where the momentum distribution is infinitely elongated along one direction, is considered. The energy loss is strongly time and direction dependent and its magnitude can much exceed the equilibrium value. It is logarithmically ultraviolet divergent.  We argue that a good approximation to the energy loss can be obtained if this divergence is cut off with the parton energy.}

\maketitle

\section{Introduction}

When a highly energetic parton travels through a quark-gluon plasma (QGP), it losses its energy due to, in particular, elastic interactions with plasma constituents. This is called {\em collisional energy loss} which for equilibrium QGP is well understood \cite{lebellac}. The quark-gluon plasma produced in relativistic heavy-ion collisions, however, reaches a state of local equilibrium only after a short but finite time interval, and during this period the momentum distribution of plasma partons is anisotropic. Consequently the system is unstable (for a review see \cite{Mrowczynski:2005ki}). 

We have developed an approach, see \cite{Carrington:2011uj,Carrington:2013tz} for a preliminary account, where energy loss is studied as an initial value problem. The approach is applicable to plasma systems evolving quickly in time. We compute the energy loss by treating the parton as an energetic classical particle with color charge. For an equilibrium plasma the known result is recovered and for an unstable plasma the energy loss is shown to have contributions which exponentially grow in time. We discuss here an extremely prolate quark-gluon plasma with momentum distribution infinitely elongated in the beam direction. Such a system is unstable due to transverse chromomagnetic modes and the spectrum of collective excitations can be obtained in explicit analytic form. The system has thus nontrivial dynamics but the computation of energy loss is relatively simple.  The energy loss appears to be strongly time and directionally dependent and its magnitude can much exceed the energy loss in equilibrium plasma. The computed energy loss is, however, ultraviolet divergent. We show that the divergence is logarithmic, as in the well-understood case of equilibrium plasma, and we argue that a good approximation to the energy loss can be obtained if this divergence is cut off with the parton energy.

\section{Formalism}
\label{sec-formalism}

Using the Wong equations \cite{Wong:1970fu}, which describe the motion of a classical parton in a chromodynamic field, one writes down the parton's energy as
\be
\label{e-loss-1}
\frac{dE(t)}{dt} = g Q^a {\bf E}_a(t,{\bf r}(t)) \cdot {\bf v} ,
\ee
where $g$ is the QCD coupling constant, $gQ^a$ is the parton color charge, ${\bf E}_a(t,{\bf r}(t))$ is  the chromoelectric field along the parton's trajectory and ${\bf v}$ is the parton's velocity. Since we consider a highly energetic parton, its velocity is assumed to be constant and ${\bf v}^2 =1$. However, neither parton's energy, nor its momentum are constant.

As we deal with an initial value problem, we apply to the field not the usual Fourier transformation but the {\it one-sided Fourier transformation} defined as
\ba
\label{1side}
f(\omega,{\bf k}) &=& \int_0^\infty dt \int d^3r
e^{i(\omega t - {\bf k}\cdot {\bf r})}
f(t,{\bf r}) , \\
\label{1side-inv}
f(t,{\bf r}) &=& \int_{-\infty+i\sigma}^{{\infty+i\sigma}} \frac{d\omega}{2\pi} \int \frac{d^3k}{(2\pi)^3}
e^{-i(\omega t - {\bf k}\cdot {\bf r})}
f(\omega,{\bf k}) ,
\ea
where the real parameter $\sigma > 0$ is chosen is such a way that the integral over $\omega$ is taken along a straight line in the complex $\omega-$plane, parallel to the real axis, above all singularities of $f(\omega,{\bf k})$. Introducing the current generated by the parton ${\bf j}_a(t,{\bf r}) = g Q^a {\bf v} \delta^{(3)}({\bf r} - {\bf v}t)$, Eq.~(\ref{e-loss-1}) can be rewritten as
\be
\label{e-loss-3}
\frac{dE(t)}{dt} = g Q^a
\int_{-\infty +i\sigma}^{\infty +i\sigma}
{d\omega \over 2\pi}
\int {d^3k \over (2\pi)^3}
e^{-i(\omega - \bar\omega)t} \; {\bf E}_a(\omega,{\bf k}) \cdot {\bf v} ,
\ee
where $\bar\omega \equiv {\bf k} \cdot {\bf v}$. 

The next step is to compute the chromoelectric field ${\bf E}_a$. Applying the one-sided Fourier transformation to the linearized Yang-Mills equations, which represent QCD in the Hard Loop approximation, we get 
\be
\label{E-field-k}
E^i_a(\omega, {\bf k}) = -i
(\Sigma^{-1})^{ij}(\omega,{\bf k})
\Big[ \omega j_a^j(\omega,{\bf k})
+ \epsilon^{jkl} k^k B_{0a}^l ({\bf k})
- \omega D_{0a}^j ({\bf k}) \Big] ,
\ee
where  ${\bf B}_0$ and ${\bf D}_0$ are the initial values of the chromomagnetic field and  the chromoelectric induction, and $D^i_a(\omega, {\bf k}) = \varepsilon^{ij}(\omega, {\bf k}) E^j_a(\omega, {\bf k})$ with  $\varepsilon^{ij}(\omega, {\bf k})$ being chromodielectric tensor;  the matrix $\Sigma$, which is the inverse retarded gluon propagator in the temporal axial gauge, is defined as
\be
\Sigma^{ij}(\omega,{\bf k}) \equiv - {\bf k}^2 \delta^{ij} + k^ik^j + \omega^2 \varepsilon^{ij}(\omega,{\bf k}) .
\ee
The equation ${\rm det}[\Sigma (\omega,{\bf k})] =0$ gives the spectrum of collective excitations.

Substituting the expression (\ref{E-field-k}) into Eq.~(\ref{e-loss-3}), we get the formula
\ba
\label{e-loss-6}
\frac{dE(t)}{dt} &=& g Q^a v^i \int_{-\infty +i\sigma}^{\infty +i\sigma}
{d\omega \over 2\pi i}
\int {d^3k \over (2\pi)^3}
e^{-i(\omega -\bar{\omega})t}
(\Sigma^{-1})^{ij}(\omega,{\bf k})
\\ [2mm]\nn
&&~~~~~~~~~~~~~~~~~~~~~~~~~~~~~~\times
\Big[ 
\frac{i \omega g Q^a v^j}{\omega - \bar{\omega}}
+ \epsilon^{jkl} k^k B_{0a}^l ({\bf k})
- \omega D_{0a}^j ({\bf k}) \Big] .
\ea
The integral over $\omega$ is controlled by the poles of the matrix $\Sigma^{-1}(\omega,{\bf k})$  
which represent the collective modes of the system. When the plasma is unstable, there are poles in the upper half-plane of complex $\omega$, and the contributions to the energy loss from these poles grow exponentially in time. Using the linearized Yang-Mills equations, the initial values ${\bf B}_0$ and ${\bf D}_0$ are expressed through the current and one obtains
\ba
\label{e-loss-unstable}
&& \frac{dE(t)}{dt} = g^2 C_R 
v^i v^l \int_{-\infty +i\sigma}^{\infty +i\sigma}
{d\omega \over 2\pi}
\int {d^3k \over (2\pi)^3}
e^{-i(\omega - \bar{\omega}) t}
(\Sigma^{-1})^{ij}(\omega,{\bf k})
\\ [2mm]\nn
&&~~~~~~~~~~~~~~~~~~~~~~~~~~~~~~\times
\Big[ 
\frac{\omega \delta^{jl}}{\omega - \bar{\omega}}
-(k^j k^k - {\bf k}^2 \delta^{jk})
(\Sigma^{-1})^{kl}(\bar{\omega},{\bf k}) 
+ \omega \, \bar{\omega} \, \varepsilon^{jk}(\bar{\omega},{\bf k})
(\Sigma^{-1})^{kl}(\bar{\omega},{\bf k})   
 \Big] \,,
\ea
which gives the energy loss of a parton in an unstable quark-gluon plasma.

When the anisotropy of the momentum distribution of plasma constituents is controlled by a single (unit) vector ${\bf n}$, it is not difficult to invert the matrix $\Sigma$. Following \cite{Romatschke:2003ms,Kobes:1990dc}, we introduce the vector ${\bf n}_T$ defined as
\be
n_T^i \equiv \big(\delta^{ij} - \frac{k^i k^j}{{\bf k}^2}\big) \, n^j
\ee 
and we use the basis of four symmetric tensors
\be
A^{ij}({\bf k}) = \delta^{ij} - \frac{k^i k^j}{{\bf k}^2} ,
\;\;\;\;\;\;
B^{ij}({\bf k}) = \frac{k^i k^j}{{\bf k}^2} ,
\;\;\;\;\;\;
C^{ij}({\bf k},{\bf n}) = \frac{n_T^i n_T^j}{{\bf n}_T^2} ,
\;\;\;\;\;\;
D^{ij}({\bf k},{\bf n}) = 
k^i n_T^j + k^j n_T^i .
\ee
Since the matrix $\Sigma$ is symmetric, it can be decomposed as $\Sigma = a\,A +b\,B +c\,C +d\,D$,  where the coefficients $a$, $b$, $c$ and $d$ are found from the equations
\ba
\label{a-b-c-d}
k^i \Sigma^{ij} k^j = {\bf k}^2 b , \;\;\;\;\;
n_T^i \Sigma^{ij} n_T^j = {\bf n}_T^2 (a + c) , \;\;\;\;\;
n_T^i \Sigma^{ij} k^j = {\bf n}_T^2{\bf k}^2 d , \;\;\;\;\;
{\rm Tr}\Sigma = 2a + b + c .
\ea
The inverse matrix, which is found from the equation  $\Sigma \,\Sigma^{-1}=1$, reads
\be
\Sigma^{-1} =
\frac{1}{a} \,A 
+ \frac{-a(a+c)\,B 
+ (- d^2{\bf k}^2{\bf n}_T^2 +bc)\,C
+ad \,D}
{a(d^2{\bf k}^2{\bf n}_T^2-b(a+c))} .
\ee
The poles of the matrix $\Sigma^{-1}$ are given by the dispersion equations
\be
\label{dis-equations}
a=0 \,, \;\;\;\;\;\;\ 
d^2{\bf k}^2{\bf n}_T^2-b(a+c) = 0 .
\ee

\section{Extremely prolate plasma}

Anisotropy is a generic feature of the parton momentum distribution in heavy-ion collisions. At the early stage, when partons emerge from the incoming nucleons, the momentum distribution is strongly elongated along the beam - it is {\em prolate} - with $\langle p_T^2 \rangle \ll \langle p_L^2 \rangle$. Due to free streaming, see {\it e.g.} \cite{Jas:2007rw}, the distribution evolves in the local rest frame to a form which is squeezed along the beam - it is {\em oblate} with $\langle p_T^2 \rangle \gg \langle p_L^2 \rangle$. We consider here the extremely prolate momentum distribution $f({\bf p}) \sim \delta(p_T)$. All quantities are expressed through the parameter analogous to the Debye mass
\be
\label{Debye-mass}
\mu^2 \equiv g^2 \int {d^3p \over (2\pi)^3} \, \frac{f({\bf p})}{|{\bf p}|},
\ee
which is the only dimensional quantity in the problem. 

\begin{figure}[t]
\centering
\includegraphics[width=0.7\textwidth]{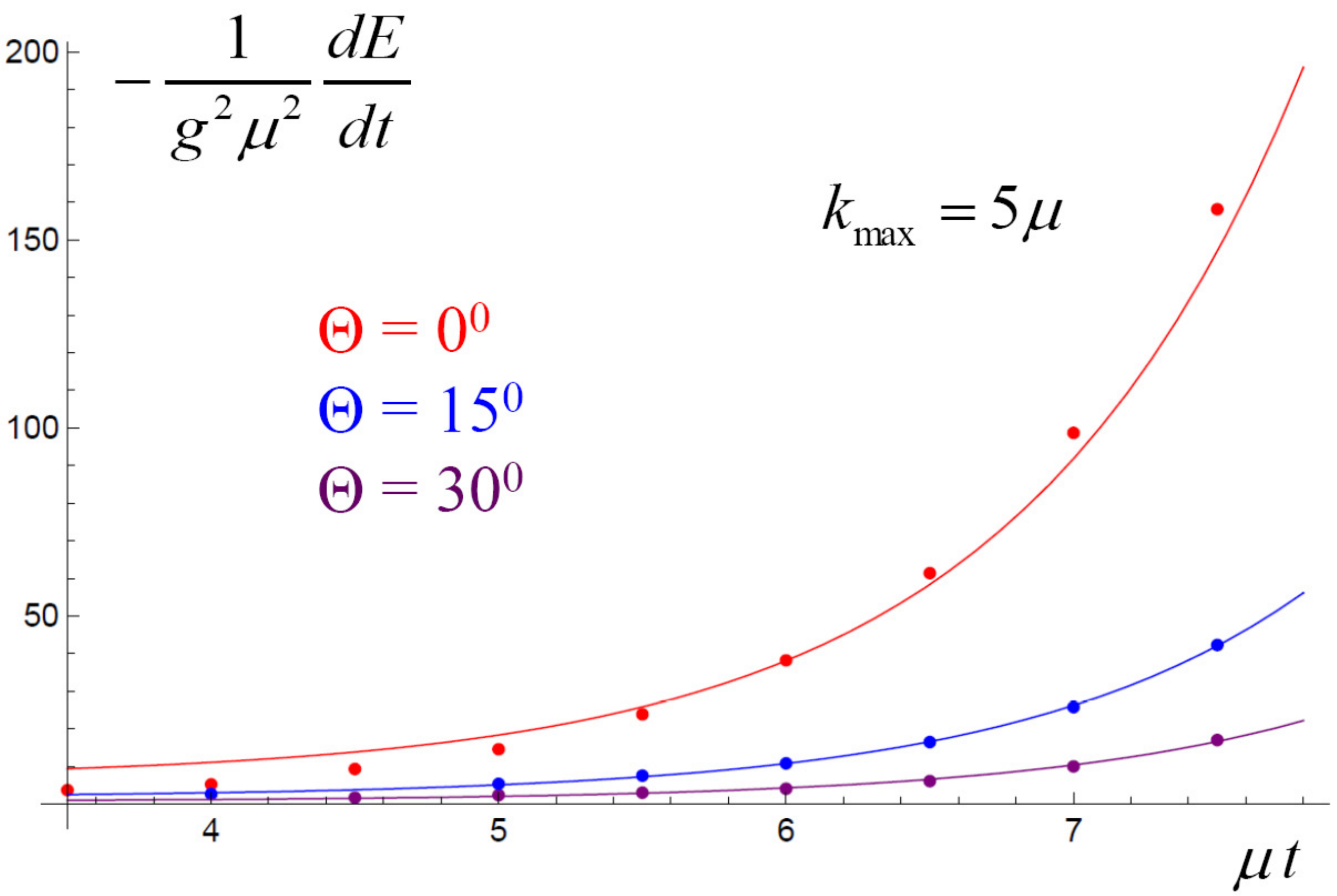}
\caption{The parton energy loss per unit time as a function of time for three angles $\Theta$ between the parton's velocity ${\bf v}$  and the axis $z$. The solid lines represent the exponential fits to the computed points.}
\label{fig-prolate-1}
\end{figure}

With the extremely prolate distribution, the spectrum of collective excitations is
\be
\label{general-solution-1}
\omega_1^2({\bf k}) = \frac{1}{2}\mu^2  + ({\bf k}\cdot {\bf n})^2 \;, 
\;\;\;\;\;\;\;\; \omega_2^2({\bf k}) = \frac{1}{2}\mu^2  + {\bf k}^2 ,
\ee
\be
\label{general-solution-2}
\omega_{\pm}^2({\bf k}) =
\frac{1}{2}\Big({\bf k}^2 + ({\bf k}\cdot {\bf n})^2
\pm \sqrt{{\bf k}^4 + ({\bf k}\cdot {\bf n})^4 
+ 2\mu^2 {\bf k}^2 - 2 \mu^2 ({\bf k}\cdot {\bf n})^2
-2 {\bf k}^2({\bf k}\cdot {\bf n})^2} \; \Big).
\ee
The modes $\omega_1$, $\omega_2$ and  $\omega_+$, are always stable. The solution $\omega_-^2$ is negative when $\mu^2 {\bf k}^2 > \mu^2 ({\bf k}\cdot {\bf n})^2+ {\bf k}^2({\bf k}\cdot {\bf n})^2$. Writing $\omega_-^2=-\gamma^2$ with $0< \gamma \in \mathbb{R}$, the solutions are $\pm i\gamma$. The first is the Weibel unstable mode and the second is an overdamped mode. 

The energy loss in the extremely prolate system is controlled by the double pole at $\omega=0$ and 8 single poles: $\omega = \pm \omega_1$, $\omega = \pm \omega_2$,  $\omega = \pm \omega_+$,  $\omega = \pm \omega_-$.  Since the collective modes are known analytically, the integral over $\omega$ is computed analytically as well. The remaining integral over ${\bf k}$ is computed numerically using cylindrical coordinates with the $z$ axis  along the vector ${\bf n}$. Since the integral is divergent, we choose a finite domain such that  $-k_{\rm max} \le k_L \le k_{\rm max}$ and  $0 \le k_T \le k_{\rm max}$ with $k_{\rm max} = 5 \mu$.  The values of remaining parameters are: $g=1$, $C_R = N_c= 3$. In Fig.~\ref{fig-prolate-1} we show the parton's energy loss per unit time as a function of time for three different angles $\Theta$ of the parton's velocity ${\bf v}$ with respect to the $z$ axis. The energy loss manifests a strong directional dependence and it exponentially grows in time, indicating the effect of instabilities. After a sufficiently long time, the magnitude of energy loss much exceeds that in equilibrium plasma which equals $0.18 \, g^2\mu^2$ for $k_{\rm max} = 5 \mu$.

\begin{figure}[t]
\centering
\includegraphics[width=0.7\textwidth]{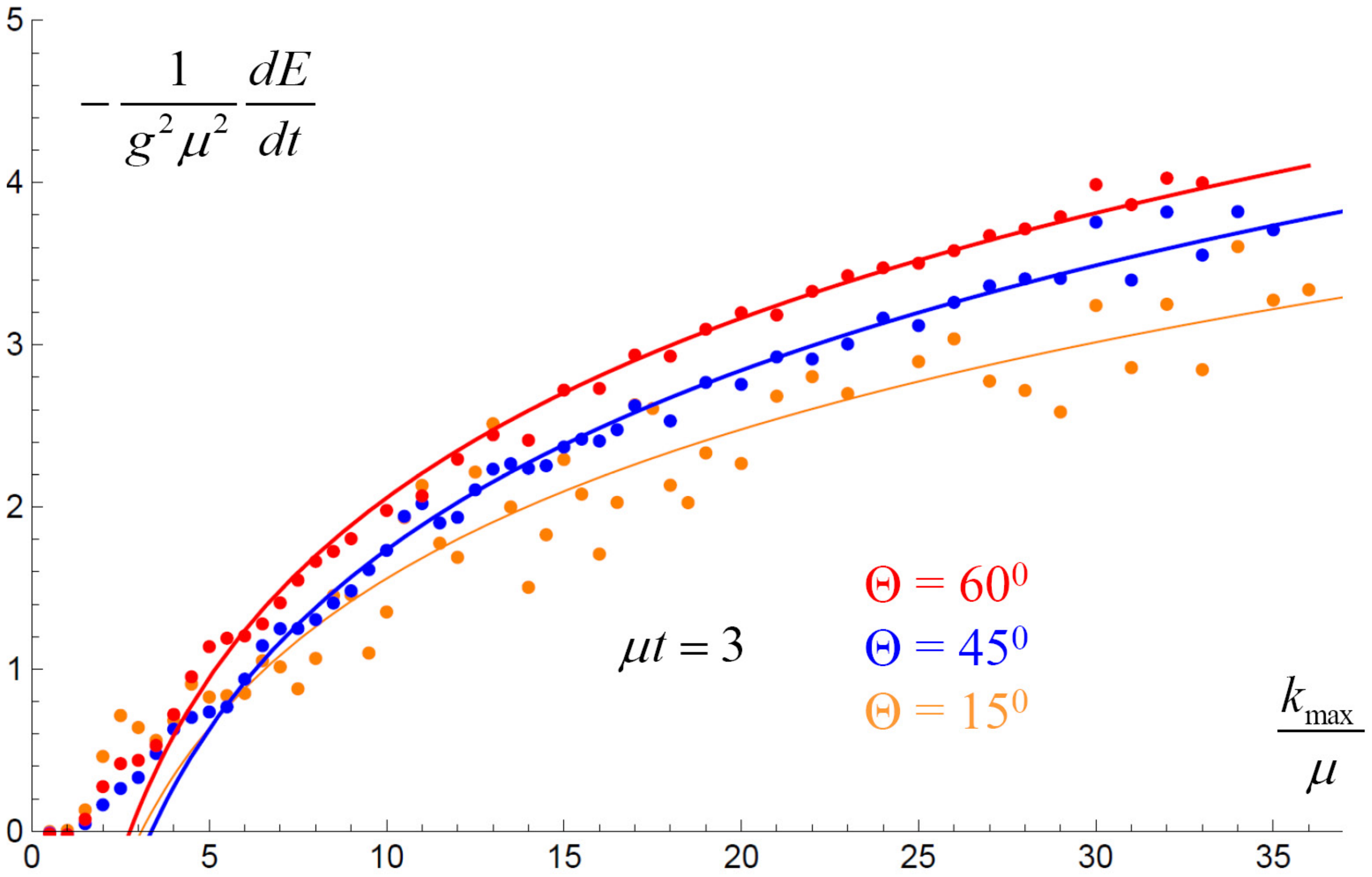}
\caption{The parton energy loss per unit time as a function of upper cut-off $k_{\rm max}$ at $t= 3 \, \mu^{-1}$  for $\Theta =  15^0$ (yellow), $\Theta =  45^0$ (blue) and $\Theta =  60^0$ (red).  The solid lines represent the logarithmic fit to the computed points.}
\label{fig-prolate-2}
\end{figure}

\section{Upper cut-off dependence}
\label{sec-cut-off}

As already mentioned, the energy loss given by the formula (\ref{e-loss-unstable}) is ultraviolet divergent. This is not surprising, as the equilibrium energy loss is divergent as well. This happens because of a breakdown of the classical approach, which is applied to the soft contribution to the energy loss, when the momentum transfer becomes too large. Consequently, one has to combine the soft contribution with the hard one, which is computed quantum-mechanically from 2 $\leftrightarrow$ 2 elastic scattering processes  \cite{lebellac}.  In equilibrium, the soft contribution to the energy loss depends logarithmically on the upper cut-off $k_{\rm max}$ divided by the Debye mass $\mu$, while the hard contribution has a logarithmic dependence on the energy of the parton $E$ divided by the same cut-off $k_{\rm max}$. The energy loss thus equals
\be
\frac{dE}{dt} = X \ln\bigg(\frac{k_{\rm max}}{\mu}\bigg) + Y  \ln\bigg(\frac{E}{k_{\rm max}}\bigg) .
\ee
In equilibrium one shows \cite{lebellac} that the coefficients $X,Y$ are equal to each other and therefore
\be
\frac{dE}{dt} = X \ln\bigg(\frac{E}{\mu}\bigg) .
\ee
The result is that the cut-offs cancel and one obtains a good approximation to the energy loss from the soft contribution with the parton energy used as an upper cut-off.

\begin{figure}[t]
\centering
\includegraphics[width=0.75\textwidth]{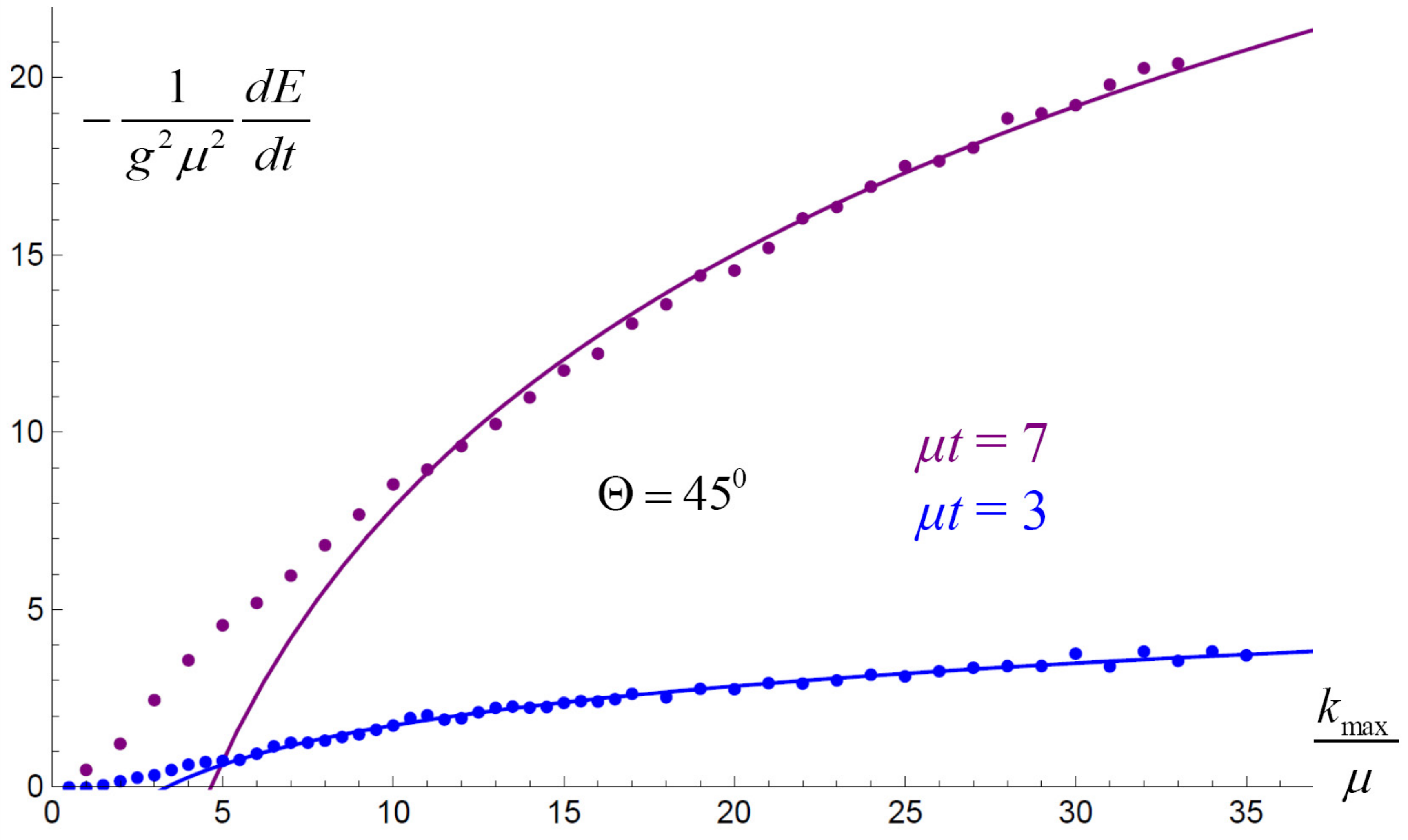}
\caption{The parton energy loss per unit time as a function of upper cut-off $k_{\rm max}$ at $\Theta =  45^0$  for $t= 3 \, \mu^{-1}$ (blue) and $t= 7 \, \mu^{-1}$ (magenta).  The solid lines represent the logarithmic fit to the computed points.}
\label{fig-prolate-3}
\end{figure}

In order to determine if this procedure is reasonable for the anisotropic case, we need to study the dependence of the energy loss in Eq.~(\ref{e-loss-unstable})  on the upper cut-off. In  Fig.~\ref{fig-prolate-2} we show the parton's energy loss per unit time as a function of $k_{\rm max}$ at $t = 3 \mu^{-1}$ for $\Theta =  15^0$, $\Theta =  45^0$ and $\Theta =  60^0$. As in the equilibrium case, the dependence is seen to be logarithmic for sufficiently large $k_{\rm max}$. However, the logarithmic dependence is modulated by oscillations which grow as the angle  $\Theta$  decreases.  Fig.~\ref{fig-prolate-3} presents the energy loss at $\Theta =  45^0$ for $t = 3 \mu^{-1}$ and $t = 7 \mu^{-1}$. For later times the onset of logarithmic dependence is observed at larger $k_{\rm max}$. Because of the logarithmic dependence, we expect to get a good approximation of the energy loss by replacing the cut-off with the parton energy.

\section{Conclusions}

We have developed a formalism where the energy loss of a fast parton in a plasma medium is found as the solution of an initial value problem.  The formalism, which allows one to obtain the energy loss in an unstable plasma, is applied to an extremely prolate quark-gluon plasma with momentum distribution infinitely elongated in the $z$ direction. This system is unstable due to chromomagnetic transverse modes. The energy loss per unit length of a highly energetic parton exponentially grows in time and exhibits a strong directional dependence.  The magnitude of the energy loss can much exceed the equilibrium value. The energy loss is ultraviolet divergent but the divergence is logarithmic. The energy loss can be obtained if this divergence is cut off with the parton energy.

\vspace{1cm}

This work was partially supported by the Polish National Science Centre under grant 2011/03/B/ST2/00110.

\end{document}